# Collimated directional emission from a peanut-shaped microresonator

Fang-Jie Shu[1,2*], Chang-Ling Zou[3], Fang-Wen Sun[3], and Yun-Feng Xiao[2*]

[1]*Department of Physics, Shangqiu Normal University, Shangqiu 476000, P. R. China*

[2]*State Key Lab for Mesoscopic Physics and Department of Physics, Peking University, Beijing 100871, P. R. China*

[3]*Key Laboratory of Quantum Information, University of Science and Technology of China, Hefei, Anhui 230026, P. R. China*

**Abstract:**

Collimated directional emission is essentially required an asymmetric resonant cavity. In this paper, we theoretically investigate a type of peanut-shaped microcavity which can support highly directional emission with the emission divergence as small as $2.5^{o}$. The mechanism of the collimated emission is explained with the short-term ray trajectory and the intuitive lens model in detail. Wave simulation also confirms these results. This extremely narrow divergence of the emission holds a great potential in highly collimated lasing from on-chip microcavities.

PACS numbers: 05.45.Mt, 42.25.-p, 42.60.Da

## I. Introduction

Whispering-gallery modes (WGMs) in microresonator systems with rotational symmetry are of current interest owing to their high Q values and small mode volumes at optical frequencies. WGMs are considered the most promising candidates for a large variety of optical applications, ranging from ultralow-threshold lasing, sensing, to cavity quantum electrodynamics [1]. One of important drawback of WGMs is their isotropic emissions due to the high degree of inherent symmetry. This causes a significant difficulty to efficiently extract and collect the microcavity emission for practical purposes. A natural choice is to design the geometrical shape of microresonators, producing a strongly directional distribution of output fields instead of isotropic emissions. These resonators are known as asymmetric resonant cavities (ARCs) or deformed cavities. Actually, shortly after the first fabrication of microdisk it was demonstrated that deforming the boundary allows for improved directionality of emission [2-4]. Since then, ARCs with directional emissions have been demonstrated in various systems including quadrupolar microdisks [4-7], full-chaos microstadiums [8-14], spiral-shape micropillars [15] or microdisks [16], liquid jet of ethanol with a quadrupole cross section [17-19], limaçon shaped microdisks [20-22], and three-dimensional deformed microspheres [23-27].

---

[*] Email address: shufangjie@gmail.com
[*] Email address: yfxiao@pku.edu.cn

Output beam divergence is an important property for ARCs, because it determines not only a high-brightness output but also a high coupling efficiency. The divergence can be estimated from the full width at half maximum (FWHM) of emission peaks in the far-field patterns (FFPs). In most ARCs [4, 6, 10, 14-17, 19, 24-27], the divergence angle typically ranges from $10^o$ to $30^o$. The divergence angle exceeds $30^o$ for limaçon shaped microdisks[20-22]. Only a special microstadium reported in Ref. [12] gets the divergence angle with several degree. To obtain a minimized divergence, Shang et al. reported a peanut-shaped cavity in which the divergence of directional emission approached $2^o$ [28]. Therein, the authors explained experimental results with a hybrid mode (a closed loop forms with a whispering gallery obit and a two-bouncing orbit), and they credited the collimation mainly to the two-bouncing orbit. In this paper, we aim to find an intuitive picture to reveal the principle that achieves the collimated emission from the peanut shaped microcavity. In Section II, we provide a brief description of the shape setting and the Poincaré surface of section (SOS) of the peanut shaped microcavity. In Section III, from a ray dynamics model, we obtain the FFPs of different peanut shapes, and then focus our attention on ray trajectory of a typical peanut-shaped cavity with the narrowest divergence. Importantly, by resorting to a new lens model, we explain the collimation emission from the peanut-shaped microcavity. In Section IV, we employ a wave method to find the double pentagons modes and confirm the ray result. Finally, in Section V, we discuss how to optimize the ideal refractive index of the microcavity material.

**II. Geometry of the peanut-shaped microcavity**

The geometry of the peanut-shaped microcavity (two-dimensional) is shown in Fig. 1(a) (shadow part). It consists of two contacted identical microdisks $C_{1,2}$ with radius of $r_1$, and a central region between them. The boundary of the central region is defined by the other two identical microdisks $C_{3,4}$ with radius of $r_2$ which are tangential to $C_{1,2}$. The whole boundary of the cavity is continuous and smooth, and looks like a peanut, thus named as a peanut-shaped microcavity.

In this design, the angle $\beta=\arccos(r_1/(r_1+r_2))$ can describe the geometry of the peanut-shaped microcavity. In our peanut shape, $\beta$ is the morphological parameter ranging from 0 to 90 degrees. When $\beta=0^\circ$, the peanut-shaped cavity is reduced to double disks in contact, which is also called photonic molecule studied extensively recently [29-31]. When $\beta=90^\circ$, the peanut-shaped cavity becomes a microstadium. For $0^\circ<\beta<90^\circ$, it is a general peanut shape studied in this paper. It should be noted that the present shaped microcavity is made of the same material with uniform refraction index $n$, which is slightly different from Ref. [28] where two silica cylinders are coated by a hybrid glass material. In Fig. 1(a), $s$ stands for the curvilinear coordinate along the boundary, and $\varphi$ denotes the far field angle measured from the main axis of the peanut.

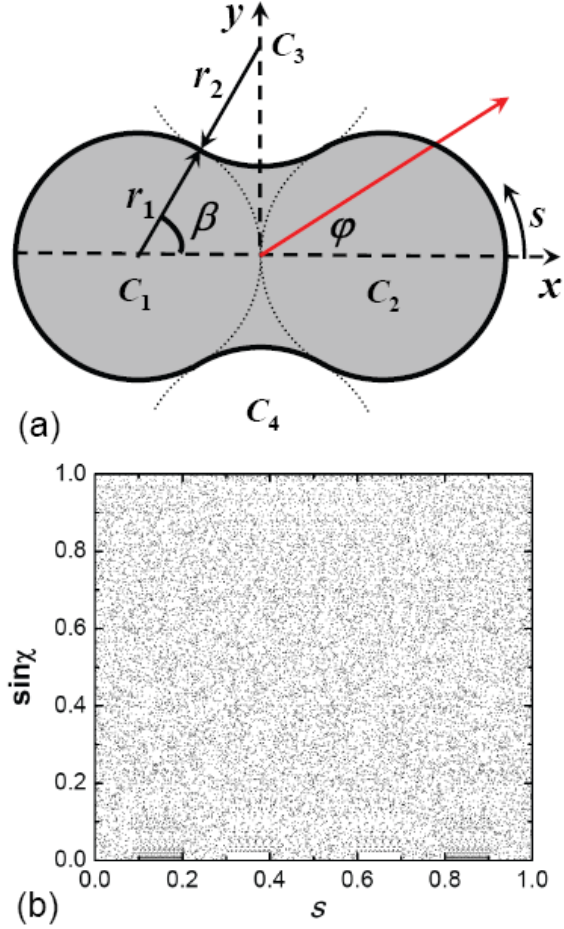

Fig. 1. (a) Peanut-shaped microcavity (shadow). Here, $s$ stands for the curvilinear coordinate along the boundary, $\beta$ describes the shape setting angle, and $\varphi$ denotes the far-field angle measured from the main axis of the peanut. (b) The SOS of the closed peanut-shaped microcavity with $\beta=60°$.

## III. Ray dynamics in the peanut-shaped microcavity

Ray dynamics provides an intuitive and efficient tool to understand the emission properties of a deformed microcavity. Thus, we first calculate the SOS of the closed peanut-shaped microcavity (the billiard) with $\beta=60°$ as plotted in Fig. 1(b). To obtain the SOS, forty rays with different initial conditions reflect on the microcavity boundary, where the ray tracing is recorded by the coordinate of reflection point ($s$, $|\sin\chi|$). Here $\chi$ represents the angle of incidence. It can be found that the ray dynamics in the SOS is fully chaotic, indicating no stable trajectory for light existed in the peanut cavity.

In the ray optic model, the ray dynamics depends on the incident angle $\chi$. When $\chi$ is larger than the critical refraction angle $\chi_c = \arcsin(1/n)$, the ray undergoes total internal reflection; otherwise, the ray splits into a reflective and a refractive rays, and the intensity of the each part is decided by well-known Fresnel's law [32]. For simplicity but without loss of the generality, in this paper we concentrate on the transverse magnetic (TM) polarized modes, whose electric field and the corresponding normal derivative are continuous crossing the boundary. The refraction intensity coefficient $T$, determined by Fresnel's law, is calculated by $1-[\sin(\chi-\chi_t)/\sin(\chi+\chi_t)]^2$, where $\chi_t$ stands for the angle of refraction given by Snell's law $n\sin\chi = \sin\chi_t$, assuming the microcavity is in the air. In the case of the closed cavity, the ray trajectory can visit the entire phase space after an enough time. However, as described above, all rays will actually refract out of the microcavity in a limited time, i.e., open cavity. Thus, we can collect the refraction rays and obtain the far-field intensity (namely, emission) distribution as a function of the far-field angle $\varphi$.

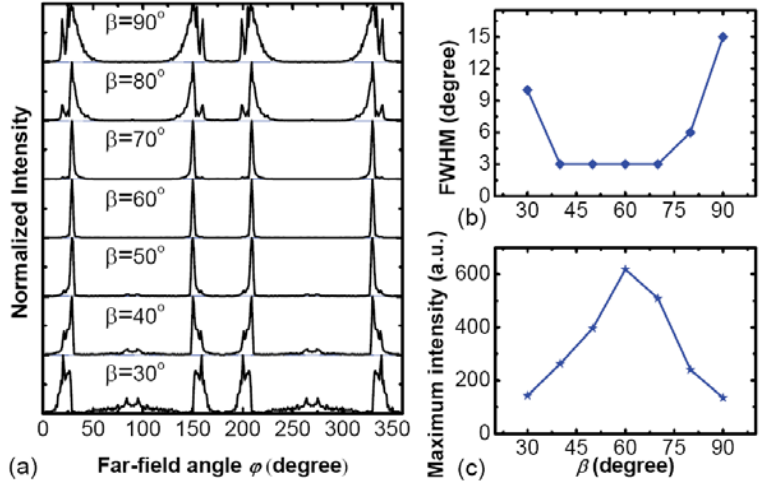

Fig. 2. (a) Normalized far-field emission patterns in the cases of different shape setting angle $\beta$. (b)-(c) FWHM, i.e., divergence angle, and maximum far-field intensity of one emission peak vs. $\beta$.

As we are interested in the relatively high-Q modes of peanut-shaped microcavities, the initial ensemble of rays is chosen to be uniformly spread in the top area of the phase space (0.93<|sin$\chi$|<1). To be consistent with Ref. [28], the refraction index of the present microcavity *n* is given as 1.52. For different *n*, it will be discussed in Section V. The similar FFPs with different shape setting angle *β* are shown in Fig. 2(a). It can be found that the peanut-shaped microcavity supports four highly directional emission angles around $\varphi$=29°, 151°, 209°, and 331°, in spite of the full chaos in SOS. It is of importance that the divergence angle, defined as the FWHM of the peak in FFP, ranges from 2.5° to 15° (see Fig. 2(b)) approximately. Evidently, Figs. 2(b) and 2(c) indicate that there exist an optimized *β*~60° ($r_1$~$r_2$) to achieve both the minimized divergence and the maximum far-field intensity. This case is named as the *regular peanut-shaped microcavity*. It is noted that, for the regular peanut, the main features of FFP agrees well with the experiment results [28]. In the following, we focus our attention on the microcavity with a regular peanut shape.

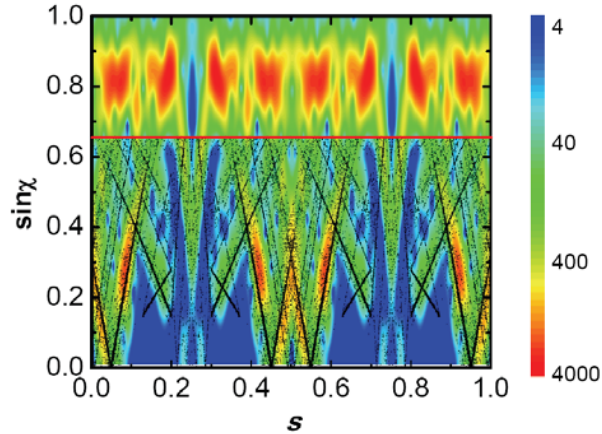

Fig. 3. Black dots: Ray simulations of short-term dynamics for random initial conditions below the critical refraction line (sin$\chi_c$=1/*n*), propagated for 60 iterations. Rainbow color: Husimi distribution for a double-pentagons mode (shown in Fig. 5(a)) projected onto the SOS of the deformed microcavity. Here, *β*=60°.

To explain this directional emission that occurs at certain angles, Fig. 3 plots unstable manifolds (black dots) of the SOS below the critical refraction line ($\chi=\chi_c$). In these manifolds, four noticeable "V"-type lines comprised of dense black dots are present, which indicates rays tend to refract out of the microcavity along them. It is of importance that the positions of them match well with the four main peaks in the FFP shown in Fig. 2(a).

To further understand collimated emission of the peanut-shape microcavity, we now employ the lens model and study the emission in both real and phase spaces. Due to the fourfold symmetry of the peanut-shaped microcavity, here we only discuss rays that leak out at the boundary of the forth quadrant, and correspondingly, we focus on the emission peak in the fourth quadrant. In this regard, the boundary arc in the second quadrant can be considered as the light source, while the right circle of the peanut plays the role of a single spherical lens which can converge lights from a point source on its focal plane into a collimated beam.

Figure 4(a) plots fifty rays in real space that totally internally reflect from the second to fourth quadrants because of the nature of concaveness in the central part. These rays finally refract out of the microcavity from the fourth quadrant as the incident angles are smaller than the critical angle. Finally, an evident collimated beam can be obtained. It should be noted that these fifty rays are randomly chosen from 2000 rays with initial ray motion shown in Fig. 4(b) (yellow dots above the critical refraction line). Before entering to the fourth quadrant, the motion of these 2000 rays finally hitting the second-quadrant boundary is also shown in the inset. It is found that these 2000 rays look likely emitted from a point light source before they reflect to the fourth quadrant and finally refract out of the microcavity. As the approximate point source is on the focal plane of the spherical lens, the refracted light behaves collimated. This result is similar to the scatter-induced directional emission in Refs. [33, 34].

Moreover, we analytically obtain all emission points in the SOS, which correspond to the same far-field emission at $\varphi=331^{\circ}$ [11], as shown in Fig. 4(b) (dashed blue curve below the critical refraction line). Importantly, this curve is finely closed to the unstable manifold (black dots), predicting a greatly collimated emission around $\varphi=331°$. In addition, although not shown here, the other three "V" shape manifolds are also closed to the corresponding curves of emission angles from the other three quadrants.

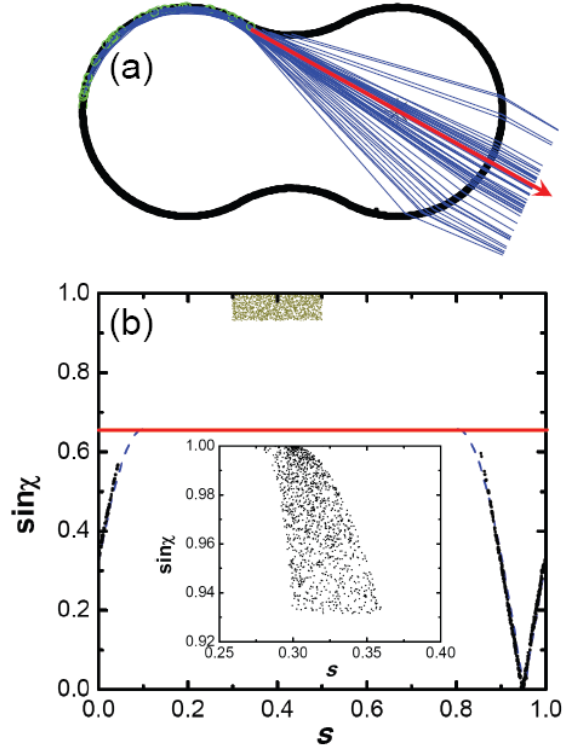

Fig. 4. (a) Fifty rays initially start in the second quadrant boundary with initial positions marked as small green circles, then reflect into the fourth quadrant, and finally refract out of the microcavity from the fourth quadrant. The red thick arrow plots the emission direction at $\varphi=331°$. (b) Yellow dots above the critical refraction line: initial motions in SOS of 2000 rays; black dots below the critical line: the unstable manifold; dashed blue curve below the critical line: all emission points in the SOS corresponding to the same far-field emission at $\varphi=331°$. Here, $\beta=60°$.

## IV. Wave correspondence

In this section, by using the boundary element method [35-37], we obtain all resonances in the range $119<nkR<121$. Therein, a high-Q resonance is excited as a lasing mode with a low threshold in practice. The highest Q approaches one thousand, and the corresponding field distribution is shown in Fig. 5 (a). Not like the preliminary interpretation in the experimental literature [28], here we find that the resonance is a double pentagons mode (black line is guided to the eye in Fig. 5(a)). Though the mode concentrates to the boundary and displays properties associated with a typical WGM, four evident emission paths cross the center parts of the two peanut kernels. This fact is different from the tangential escape in the slightly deformed microcavities [25]. The FFP of this double pentagons mode is also shown in Fig. 5(b). It agrees well with both the ray trajectory simulation in Fig. 2(a) and the experiment result in Ref. [28].

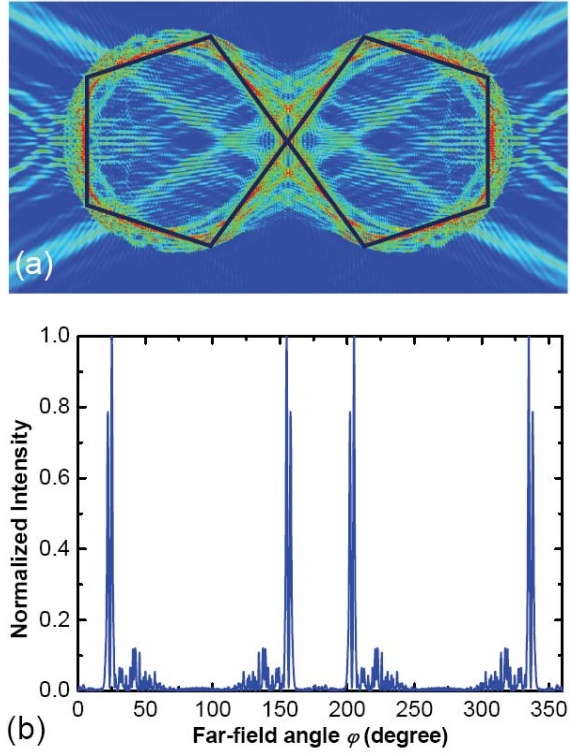

Fig. 5. (a) Near-field pattern of double pentagons mode in real space. The black line marks the eye-guided periodic orbit. (b) The far-field emission pattern for the peanut-shaped cavity with $n=1.52$ and $\beta=60°$.

To further study the properties of this double pentagons mode, now we perform the Husimi projection [38], which represents the wave analog of the SOS. The logarithm intensity distribution of the Husimi projection is shown in Fig. 3(a) (rainbow color). In the region above the critical refraction line ($\chi=\chi_c$, red line), eight scars at appropriate positions do exist, demonstrating the double pentagons mode even in the fully chaotic peanut-shaped microcavity.

In the leaky region, the Husimi projection is also in good agreement with the unstable manifold in detail.

**V. Discussions**

As demonstrated above, the peanut-shaped microcavity owns the merit of high collimation. Here we provide some brief discussions to improve its performances for extensive use. In general, the output performance of deformed microcavity strongly depends on both the cavity geometry and the refraction index of material. In Section II we have defined one morphology parameter $\beta$. Actually, if the two circles $C_{1,2}$ are not contacted, the gap between them becomes another morphology parameter. The gap also plays a significant role in the directional emission of the peanut-shaped cavity, determining the directions and divergence, similar to the case that in a stadium-shaped microcavity [12]. In addition, if the two circles are not identical, the radii ratio is another morphology parameter. In this case, the four-fold symmetry of the regular peanut-shaped microcavity is destroyed and the optical vernier effect [39] will appear for the mode which travels both kernels of the peanut.

In the discussion above, the refraction index $n$ of the cavity material is assumed as 1.52. Now we turn to study the emission property with $n$ changing. On one hand, Fig. 6(a) depicts the divergence angle (i.e., the FWHM of the emission peak) with $\beta$=45º, 60º and 75º. It is found that in our peanut-shaped microcavities, the highly collimated emission appears at low $n$ range. The minimized divergence angle produces at $n$~1.5. When $n$ is larger than 2.2, the divergence angle even exceeds 60º in the case of $\beta$=45°, because the two central peaks may be overlapped (see the inset). On the other hand, a bright far-field point is of essence. To evaluate it, we obtain the highest intensity in FFP at the different $n$, as shown in Fig. 6(b). It can be found that there has an ideal $n$ that lies in low index region for each $\beta$. Note that for the regular peanut-shaped microcavity, the ideal refraction index is about 1.52. The peak intensity approaches 633 units with the total initial intensity of 8000 units in ray trajectory. In other words, about 8% energy is concentrated in the angle of one degree. This is 28 times of isotropy emitting energy. If our receptor has ~3° flare angle, the received energy exceeds 20%.

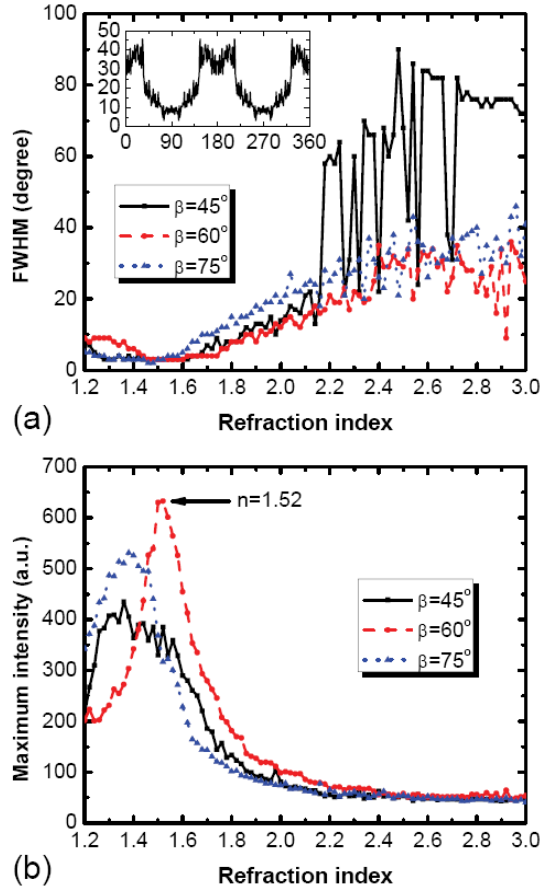

Fig. 6. FWHM of main peaks (a) and maximum far-field intensities (b) depending on the refraction index *n* with *β*=45°, 60° and 75°. The inset indicates that the two central peaks may overlap when *n* is higher than 2.2 in the case of *β*=45°.

## VI. Conclusion

In summary, we have theoretically studied the properties of directional emission of a type of peanut-shaped microcavity. The short-term dynamics in ray simulations points out that the deformed microcavity support four evident emission directions; while a lens model is employed to demonstrate the collimated emission with the divergence as small as 2.5°. The wave simulation is also provided to show the resonance pattern and Husimi projection, and double pentagons obit appears though the peanut-shaped microcavity is fully chaotic. The FFP obtaining from the wave simulation agrees well with the result given by ray and experiment. Remarkably, this obit is different from the explanation in Ref. [28]. The extremely narrow divergence of the emission holds a great potential in highly collimated lasing from on-chip microcavities.

**Acknowledgments**

This work was supported by the Foundation of He'nan Educational Committee (No. 2011A140021), the Basic and High-tech Project of He'nan Province, and the Youth Foundation of Shangqiu Normal University (No. 2010QN15). YFX acknowledges support from the National Science Foundation of China (No. 10821062 and No. 11004003), the National Basic Research Program of China (No. 2007CB307001), and the Research Fund for the Doctoral Program of Higher Education (No. 20090001120004).